%% file: manuscript_arxiv_stee1.tex
\documentclass[pra,twocolumn,superscriptaddress,showpacs]{revtex4-1}

\usepackage{amsmath}
\usepackage{latexsym}
\usepackage{amssymb}
\usepackage{graphicx}
\usepackage[colorlinks=true, citecolor=blue, urlcolor=blue]{hyperref}
\usepackage{float}
\usepackage{amsfonts}
\usepackage{textcomp}
\usepackage{mathpazo}
\usepackage{comment}
\usepackage{xr}
\usepackage{amsthm}

\newtheorem{theorem}{Theorem}

\usepackage{bbm}

\usepackage{xcolor}
\definecolor{myurlcolor}{rgb}{0,0,0.4}
\definecolor{mycitecolor}{rgb}{0,0.5,0}
\definecolor{myrefcolor}{rgb}{0.5,0,0}
\usepackage{hyperref}
\hypersetup{colorlinks,
linkcolor=myrefcolor,
citecolor=mycitecolor,
urlcolor=myurlcolor}

\usepackage[draft]{fixme}
\usepackage{amsmath,bbm}
\usepackage{graphicx}
\usepackage{amsfonts}
\usepackage{amssymb}
\usepackage{amsmath, amssymb, amsthm,verbatim,graphicx,bbm}
\usepackage{mathrsfs}
\usepackage{color,xcolor,longtable}

\usepackage{xr}
\makeatletter
\newcommand*{\addFileDependency}[1]{
  \typeout{(#1)}
  \@addtofilelist{#1}
  \IfFileExists{#1}{}{\typeout{No file #1.}}
}
\makeatother

\newcommand*{\myexternaldocument}[1]{
    \externaldocument{#1}
    \addFileDependency{#1.tex}
    \addFileDependency{#1.aux}
}

\myexternaldocument{manuscript_PRL}

\newcommand{\beq}[0]{\begin{equation}}
\newcommand{\eeq}[0]{\end{equation}}

\def\ra{\rangle}
\def\la{\langle}

\newcommand{\one}{\leavevmode\hbox{\small1\normalsize\kern-.33em1}}

\def\be{\begin{equation}}
\def\ee{\end{equation}}
\def\ben{\begin{eqnarray}}
\def\een{\end{eqnarray}}
\def\eea{\end{array}}
\def\bea{\begin{array}}

\newcommand{\Tr}[1]{\mathrm{Tr}#1}
\newcommand{\bei}{\begin{itemize}}
\newcommand{\eei}{\end{itemize}}
\newcommand{\ket}[1]{|#1\rangle}
\newcommand{\bra}[1]{\langle#1|}

\newcommand{\proj}[1]{\ket{#1}\!\bra{#1}}

\def\ra{\rangle}
\def\la{\langle}

\newcommand{\I}{\mathbbm{1}}

\renewcommand{\emph}[1]{\textbf{#1}}

\newcommand{\eqnref}[1]{(\ref{#1})}

\makeatletter
\newtheorem*{rep@theorem}{\rep@title}
\newcommand{\newreptheorem}[2]{%
\newenvironment{rep#1}[1]{%
 \def\rep@title{#2 \ref{##1}}%
 \begin{rep@theorem}}%
 {\end{rep@theorem}}}
\makeatother

\theoremstyle{plain}
\newtheorem{thm}{Theorem}
\newtheorem*{thm*}{Theorem}
\newreptheorem{thm}{Theorem}
\newtheorem{lem}{Lemma}

\newtheorem*{udefn}{Definition}
\theoremstyle{definition}

\theoremstyle{remark}

\usepackage[T1]{fontenc}

\begin{document}
\title{Certification of incompatible measurements using quantum steering}
\author{Shubhayan Sarkar}
\affiliation{Center for Theoretical Physics, Polish Academy of Sciences, Aleja Lotnik\'{o}w 32/46, 02-668 Warsaw, Poland}
\author{Debashis Saha}
\affiliation{Center for Theoretical Physics, Polish Academy of Sciences, Aleja Lotnik\'{o}w 32/46, 02-668 Warsaw, Poland}

\affiliation{School of Physics, Indian Institute of Science Education and Research Thiruvananthapuram, Kerala, India 695551 }

\author{Remigiusz Augusiak}
\affiliation{Center for Theoretical Physics, Polish Academy of Sciences, Aleja Lotnik\'{o}w 32/46, 02-668 Warsaw, Poland}

\begin{abstract}
In this letter we consider the problem of certification of quantum measurements with an arbitrary number of outcomes. We propose a simple scheme for certifying any set of $d$-outcome projective measurements which do not share any common invariant proper subspace, termed here genuinely incompatible, and the maximally entangled state of two qudits. For our purpose, we focus on a simpler scenario, termed as one-sided device-independent scenario where the resource employed for certification is quantum steering. We also study robustness of our self-testing statements for a certain class of genuinely incompatible measurements including mutually unbiased bases which are essential for several quantum information-theoretic tasks such as quantum cryptography.
\end{abstract}

\maketitle 


\textit{Introduction.---}The existence of nonlocal correlations, 
referred to as nonlocality, is one of the most intriguing features of quantum theory \cite{NonlocalityReview}. Defined for the first time in a rigorous way by Bell \cite{Bell,Bell66}, nonlocality is considered now a powerful resource for
a plethora of applications in the device-independent (DI) regime, 
with the most celebrated example being quantum cryptography \cite{DICrypto}.

Another interesting application of nonlocality is self-testing. 
Introduced in Ref. \cite{Yao}, it represents the most complete form 
of device-independent certification allowing one to almost fully 
(up to some equivalences such as local unitary operations and 
extra degrees of freedom) characterise the form of 
the quantum state and the measurements performed on it
only from the observed non-local correlations \cite{SupicReview}. Although 
recently there has been a substantial progress in designing 
self-testing schemes \cite{Scarani, Bamps, All, chainedBell, Projection,  sarkar, prakash, Jed1, Armin1}, most of them are dedicated to entangled quantum states, 
leaving the question of certification of quantum measurements
largely unexplored. In particular, while 
there exists a self-testing strategy for any bipartite 
entangled state of arbitrary local dimension \cite{Projection}, 
apart from a few classes of arbitrarily-dimensional measurements such as 
mutually unbiased measurements \cite{Armin1} (see also \cite{Jed1})
or the optimal CGLMP measurements \cite{sarkar},
no general method for DI certification of 
quantum measurements has been proposed so far. In fact,
deriving such a scheme is a hard task because there exist 
incompatible measurements which do not give rise to any nonlocality \cite{Brunner,Vertesi}.
Moreover, devising certification methods becomes 
increasingly difficult when one goes beyond low-dimensional quantum systems.



To reduce its complexity one can consider
some simplified scenarios 
in which certain physically-motivated assumptions are made about 
the devices. An example is the prepare-and-measure (PM) scenario 
in which the dimension of the underlying quantum system
is constrained. This is a strong assumption as it 
kills one of the aforementioned equivalences 
associated with the extra degrees of freedom present in the DI setting. 
Taking advantage of this scenario, certification methods for 
a couple of interesting classes of quantum measurements have been introduced 
including non-projective qubit measurements \cite{Armin2, Marcin}, 
mutually unbiased bases \cite{Jed2} and non-projective SIC measurements \cite{Armin3} 
in arbitrary dimension. Also, a strategy allowing to certify 
overlaps between any number of measurement bases has been 
proposed in \cite{Miklin}, which, nevertheless, translates
to a method of self-testing of projective measurements only 
in the simplest two-dimensional case. No such general scheme for 
arbitrarily-dimensional measurements is known in the PM scenario.

In this work we address the problem of certification
of quantum measurements in yet another relaxation of the DI setting:
the one-sided-device-independent scenario (1SDI). 
It directly derives from the Bell scenario by assuming 
that one of the measuring devices is trusted and performs 
well-known measurements and is based on 
quantum steering---another
form of quantum correlations \cite{Wiseman, Ecaval, Quin}. 
Importantly, unlike in the Bell scenario there is a one-to-one correspondence between quantum steering and measurement incompatibility \cite{Guhne,quintino, Caval1}, 
and thus this scenario is perfectly suited for designing certification schemes for quantum measurements. Furthermore, contrary to the prepare-and-measure scenario, it also allows 
for certification of entangled quantum states.

We provide a simple scheme for self-testing
of $d$-outcome incompatible projective 
measurements and the two-qudit maximally entangled state. 
Our scheme is designed for 
quantum observables which do not share a common 
invariant subspace (termed here genuinely incompatible) such
as those corresponding to mutually
unbiased bases (MUBs) in $d$-dimensional Hilbert spaces.
To this end, we introduce a general family of steering inequalities 
inspired by the constructions of Refs. \cite{Horodecki,Stein}.
We also discuss the case when the certified measurements 
are not genuinely incompatible and study robustness of our 
scheme for a particular choice of 
observables.

\textit{Preliminaries---}We first outline the scenario and introduce the relevant terminology.

\textit{Steering scenario.}
We consider a standard bipartite scenario exhibiting quantum steerability, however, 
%
%
we formulate it in a way resembling the Bell scenario. Consider two spatially separated observers Alice and Bob who share a quantum state $\rho_{AB}$ acting on $\mathcal{H}_A\otimes\mathcal{H}_B$ and perform quantum measurements on their shares of this state. Contrary to the DI setting, we additionally assume that Alice's measuring device is trusted and performs well-known quantum measurements; Bob's measurement device is untrusted. We also assume that the dimension of Alice's Hilbert space is fixed: $\dim\mathcal{H}_A=d$.

We denote Alice's and Bob's measurements
by $M_x=\{M_{a|x}\}$ and $N_y=\{N_{b|y}\}$, respectively, where $M_{a|x}\geq 0$ and $N_{b|y}\geq 0$ are the measurement operators that sum up to the identity; $x,y=1,\ldots,N$ and $a,b=0,\ldots,d-1$ denote measurement choices and outcomes, respectively.
Correlations observed in this experiment are described by 
a collection of joint probability distributions $\{p(a,b|x,y)\}$, 
where $p(a,b|x,y)=\Tr[(M_{a|x}\otimes N_{b|y})\rho_{AB}]$.

To detect that a state $\rho_{AB}$ exhibits quantum steering
one uses the following linear Bell-type inequalities
\begin{eqnarray}\label{functionals}
\mathcal{B}\left(p(a,b|x,y)\right):=\sum_{a,b,x,y}\alpha_{a,b|x,y}p(a,b|x,y) \leq \beta_L,
\end{eqnarray}
where $\beta_L$ is the maximal value of $\mathcal{B}\left(p(a,b|x,y)\right)$
over correlations admitting local hidden state models \cite{Wiseman} 
%
%
(see Appendix B of \cite{SupMat}). If violated, such an inequality implies that $\rho_{AB}$ is steerable. The maximal value of $\mathcal{B}\left(p(a,b|x,y)\right)$ achievable in quantum theory is denoted $\beta_Q$.

\textit{Incompatible measurements.} Let us also recall the notion of incompatibility of quantum measurements. Consider two projective measurements 
$M_x$ $(x=1,2)$ and the associated observables. We call $M_x$ incompatible 
if these observables do not commute. Below we introduce a more restrictive definition of incompatible measurements that not only requires them to be
noncommuting but also not to possess a common invariant subspace. 

It is convenient to 
express correlations observed in the above experiment in terms of 
expectation values instead of probability distributions. Given that here we consider $d$-outcome measurements we employ the notion of generalized correlators
(see Refs. \cite{Son2006,Liang2009}) which are complex quantities defined as
the Fourier transform of $\{p(a,b|x,y)\}$,
\begin{eqnarray}\label{ft}
\langle A_{k|x}B_{l|y}\rangle=\sum_{a,b=0}^{d-1}\omega^{ak+bl}p(a,b|x,y),
\end{eqnarray}
where $\omega=\exp(2\pi \mathbbm{i}/d)$ and $ k,l\in\{0,1,\ldots,d-1\}$. 
In the quantum case 
these generalized correlators express as
$\langle A_{k|x}B_{l|y}\rangle=
\Tr[\rho_{AB}(A_{x}^k\otimes B_{l|y})]$,
where $A_x$ are unitary operators whose eigenvalues are $\omega^i$
$(i=0,\ldots,d-1)$, and the superscripts are operator powers, whereas 
$B_{l|y}$ are Fourier transforms of Bob's measurements operators: $B_{l|y}=\sum_{b}\omega^{bl}N_{b|y}$. 
The Fourier transform is invertible and hence the set $B_y=\{B_{k|y}\}$ 
fully determines Bob's measurement $N_y$. 
Thus, we can unambiguously represent Bob's measurements by the corresponding 
sets $B_y$.
If $N_y$ is projective, then $B_{k|y}$ is 
the $k$th power of a unitary observable, which, slightly abusing the notation, 
we also denote $B_y$, i.e., $B_{k|y}=B_y^k$ (see Appendix A of \cite{SupMat} for details).

{\it{Self-testing.}} Self-testing in the steering scenario was first defined 
in Ref. \cite{Supic}. Inspired by \cite{SupicReview}, here we state a more general version that does not assume the underlying state to be pure and measurements to be projective.

Consider again the above experiment with Alice and Bob performing measurements on a state $\rho_{AB}$ and observing correlations $\{p(a,b|x,y)\}$; recall that $A_x$ are projective and fixed, whereas Bob's measurements represented by $B_y=\{B_{k|y}\}$ are arbitrary. Consider then a reference experiment involving a state $\ket{\psi'}_{AB}$ and measurements represented by $B_y'=\{B_{k|y}'\}$ that we both want to certify. These experiments are equivalent, or, alternatively, $\ket{\psi'}_{AB}$ and $B_y'$ are certified from $\{p(a,b|x,y)\}$ if there exists a unitary $U_B:\mathcal{H}_B\to \mathcal{H}_B$ such that 
\begin{equation}
 (\mathbbm{1}_A\otimes U_B)\rho_{AB}(\mathbbm{1}_A\otimes U_B^{\dagger})=\proj{\psi'}_{AB'}\otimes\rho_{B''},
\end{equation}
\begin{equation}
    U_B\,B_{k|y}\,U_B=B_{k|y}'\otimes\mathbbm{1}_{B''},
\end{equation}
where $\mathcal{H}_B$ decomposes as $\mathcal{H}_B=\mathcal{H}_{B'}\otimes\mathcal{H}_{B''}$, and $\rho_{B''}$ and $\mathbbm{1}_{B''}$ is some state and the identity acting on $\mathcal{H}_{B''}$.

{\it{Results.---}}We begin by introducing a class of inequalities for testing nonclassicality (\ref{functionals}) inspired by those considered in Refs. \cite{Horodecki,Stein}. Here we state them in the correlator picture as (see Appendix A of \cite{SupMat})
%
\begin{eqnarray}\label{SteeIn}
\mathcal{B}_{d,N}=\sum_{i=1}^{N}\sum_{k=1}^{d-1}\langle A_i^{k}\otimes B_{k|i}\rangle\leq \beta_L,
\end{eqnarray} 
where, again, $A_i$ are fixed $d$-outcome unitary observables whose eigenvalues are powers of $\omega$, whereas $B_{k|i}$ are Fourier transforms of Bob's measurement operators that act on an unknown finite-dimensional Hilbert space $\mathcal{H}_B$. 
Notice that steering inequalities in the correlator picture have already been 
considered, e.g., in Ref. \cite{Saunders2010}.

The maximal quantum value of $\mathcal{B}_{d,N}$ is
$\beta_Q=N(d-1)$ and it is achieved by the two-qudit maximally entangled state, 
\be\label{phiplus}
\ket{\phi_d^+}=\frac{1}{\sqrt{d}}\sum_{i=0}^{d-1}\ket{ii},
\ee 
and Bob's observables $B_i=A_i^*$. Then, the maximal 
classical value of $\mathcal{B}_{d,N}$ can be upper bounded as
\begin{eqnarray}\label{classBoun}
\beta_L\leq\max_{{\rho}}\sum_{i=1}^{N}\sum_{k=1}^{d-1} \left|\langle A_i^{k}\rangle_{\rho}\right|.
\end{eqnarray}
%
It follows from this bound that the steering inequality \eqref{SteeIn} is non-trivial ($\beta_L<\beta_Q$) whenever the observables $A_i$ do not share a common eigenvector. While this is sufficient 
for the inequality (\ref{SteeIn}) to exhibit quantum violation, 
it is not for deriving our self-testing statements. In fact, we need to restrict ourselves to a smaller class of quantum observables defined as follows. 
\begin{udefn}
Consider a set of $N$ $d$-outcome unitary observables $A_i$ 
acting on $\mathbbm{C}^d$ and obeying $A_i^d=\mathbbm{1}$. 
We call them genuinely incompatible (GI) if there is 
no subspace $V\subset \mathbbm{C}^d$
such that $\dim V<d$ and $A_iV\subseteq V$ for all $i$; 
%
%
in other words, the only common invariant space of all 
$A_i$ is the full space $\mathbbm{C}^d$.
\end{udefn}

An excellent example of genuinely 
incompatible (GI) observables are those whose eigenvectors form 
MUBs in $\mathbbm{C}^d$ (see Appendix C of \cite{SupMat}) 
such as for instance $X_dZ_d^k$ for prime $d$ with $k=0,\ldots,d-1$; 
here, $X_d$ and $Z_d$ are $d$-dimensional generalizations
of the qubit Pauli matrices, defined as $X_d\ket{i}=\ket{i+1}$
and $Z_d\ket{i}=\omega^i\ket{i}$, respectively, where $\ket{i}$ are
the elements of the computational basis in $\mathbbm{C}^d$ and the addition is modulo $d$.

Let us make a few observations regarding the above definition. First, genuinely incompatible observables do not share a common eigenvector, and thus give rise to a nontrivial inequality (\ref{SteeIn}). The opposite implication is in general not true: observables $A_i$ with no common eigenvector might not be genuinely incompatible. Second, any set of observables containing two GI measurements like $X_d$ and $Z_d$ is genuinely incompatible, however, there exist observables forming GI set while not being pairwise genuinely incompatible (see Appendix C of \cite{SupMat} for an example).
Third, the following fact holds true for any set of GI observables (see Appendix C of \cite{SupMat} for a proof).

\begin{lem}\label{lemma4}
Consider a set of $N$ $d$-outcome unitary observables $A_i$ such that $A_i^d=\mathbbm{1}_d$,
and assume them to be GI. The following implication holds true for any normal matrix $P$ on $\mathbbm{C}^d$: if  $[P,A_i]=0$ for all $i$, then $P=\lambda\mathbbm{1}_d$ for some $\lambda\in\mathbbm{C}$.
\end{lem}

Now, we proceed to the main results of this work.


\begin{thm}\label{Theo1M} 
Assume that for a given set of GI unitary observables $A_i$ obeying $A_i^d=\mathbbm{1}_d$, the steering inequality \eqref{SteeIn} is maximally violated by a state $\rho_{AB}$ acting on $\mathbbm{C}^d\otimes\mathcal{H}_B$ and measurements represented by $B_i$. Then, the following statements hold true for any $d$: (i) Bob's measurements are projective, (ii) Bob's Hilbert space decomposes as $\mathcal{H}_{B}=(\mathbbm{C}^d)_{B'}\otimes \mathcal{H}_{B''}$ for some Hilbert space $\mathcal{H}_{B''}$, (iii) there exists a unitary transformation, $U_B:\mathcal{H}_B\rightarrow\mathcal{H}_B$, such that
\begin{eqnarray}\label{lem1.2}
(\mathbbm{1}_A\otimes U_B)\rho_{AB}(\mathbbm{1}_A\otimes U_B^{\dagger})=\proj{\phi^+_d}_{AB'}\otimes \rho_{B''},
\end{eqnarray}
where $B''$ denotes Bob's auxiliary system, and 
\begin{eqnarray}\label{lem1.1}
\forall i, \quad U_B\,B_i\,U_B^{\dagger}=A_i^{*}\otimes \mathbbm{1}_{B''}.
\end{eqnarray}
\end{thm}

\begin{proof}Here we present a sketch of the proof
(see Appendix C of \cite{SupMat} for the full version).
The maximal quantum value $\beta_{Q}=N(d-1)$ of the inequality 
\eqref{SteeIn} equals its maximal algebraic value. 
Thus, if maximally violated it forces $\rho_{AB}$, 
$A_i$ and $B_{k|i}$ to satisfy
\begin{eqnarray}\label{SOS00}
(A_i^{k}\otimes B_{k|i}) \rho_{AB} = \rho_{AB}
\end{eqnarray}
for all $i$ and $k$. By applying $A_i^{d-k}\otimes B_{d-k|i}$
to both sides and using the fact that, by definition, $B_{d-k|i}=B_{k|i}^{\dagger}$, 
one finds that $B_{k|i}$ are all unitary, and thus each 
Bob's measurements is projective; we denote $B_{k|i}=B_i^k$.
Let us add that throughout our work we assume that Bob's reduced state $\rho_B=\Tr_A(\rho_{AB})$ is full rank \footnote{In other words, we assume that the dimensions of Bob's measurement operators are the same as those of his reduced state $\rho_B$. This assumption is fully justified because we are able to make nontrivial statements on those parts of Bob's measurement that act on $\mathrm{supp}(\rho_{B})$; what is beyond $\mathrm{supp}(\rho_{B})$ does not contribute to the observed non-classicality.}.

We then observe that the condition (\ref{SOS00}) implies 
that all pure states belonging to the support of $\rho_{AB}$
must also satisfy it, that is,
%
%
\begin{eqnarray}\label{SOS01}
(A_i^{k}\otimes B_{i}^k) \ket{\psi}_{AB} = \ket{\psi}_{AB}
\end{eqnarray}
holds true to any $i$ and $k$ and $\ket{\psi}_{AB}\in\mathrm{supp}(\rho_{AB})$. 
Next, we consider the Schmidt decomposition of $\ket{\psi}_{AB}$,
\begin{equation}\label{Schmidt}
    \ket{\psi}_{AB}=\sum_{i=0}^{d-1}\lambda_i\ket{e_i}\ket{f_i},
\end{equation}
where the coefficients $\lambda_i\geq 0$ satisfy $\sum_{i}\lambda_i^2=1$,
and the local vectors $\ket{e_i}\in\mathbbm{C}^d$ and $\ket{f_i}\in\mathcal{H}_B$ 
are orthonormal. Notice that the number of terms in this decomposition is a 
consequence of the assumption that the dimension of 
Alice's Hilbert space is at most $d$.

Let us now make a few manipulations with (\ref{Schmidt}). 
First, we observe that there exists a unitary $\mathcal{U}_B$ such that 
$\ket{f_i}=\mathcal{U}_B\ket{e_i^*}$ for every $i$, where the asterisk denotes 
complex conjugation in the standard basis; thus
\begin{eqnarray}\label{state2}
\ket{\psi}_{AB}=(P_A\otimes\mathcal{U}_B)
\frac{1}{\sqrt{d}}\sum_{i=0}^{d-1}\ket{e_i}\ket{e_i^*},
\end{eqnarray}
where $P_A=\sqrt{d}\sum_i\lambda_i\proj{e_i}$.
%
%
The fact that $A_i$ are GI implies that $P_A$ is full rank, i.e., $\lambda_i>0$ for any $i$. We finally notice that the state appearing on the right hand side of (\ref{state2}) is the maximally entangled state (\ref{phiplus});  consequently 
\begin{equation}\label{PW}
\ket{\psi}_{AB}=(P_A\otimes \mathcal{U}_B)\ket{\phi^{+}_d}.
\end{equation}

Let us also notice that Eq. (\ref{SOS00}) forces each observable 
$B_i$ to have a block structure $B_i=\overline{B}_i\oplus E_i$,
where $\overline{B}_i$ is a $d\times d$ unitary whose eigenvalues are powers of $\omega$ and acts on the support of $\Tr_A\psi_{AB}$, that is, 
the subspace spanned by the local vectors $\ket{f_i}$. Crucially, 
although the other block $E_i$ is a proper observable too, it 
trivially acts on the state $\ket{\psi}_{AB}$. 
Thus, below we focus on $\overline{B}_i$.

Combining Eq. \eqref{SOS00} for $k=1$ with Eq. \eqref{PW}, we obtain
\begin{eqnarray}
(A_iP_A\otimes \widetilde{B}_i)\ket{\phi^{+}_d}=(P_A\otimes \mathbbm{1}_B)\ket{\phi^{+}_d}\nonumber,
\end{eqnarray}
where $\widetilde{B}_i=\mathcal{U}_B\, \overline{B}_i \,
\mathcal{U}_B^\dagger$. Using then the identity $R\otimes Q\ket{\phi_d^{+}}=RQ^T\otimes\mathbbm{1}\ket{\phi_d^{+}}$
that holds for arbitrary matrices $R$ and $Q$,
and omitting the state $\ket{\phi_d^+}$, 
we arrive at a matrix equation
%
$A_iP_A\widetilde{B}_i^T=P_A$.
%
After multiplying this relation by its hermitian conjugation
and taking advantage of the facts that 
$A_i$ and $\widetilde{B}_i$ are unitary
and that $P_A\geq 0$, we can bring it
to a simple condition for $P_A$:
\begin{equation}
    [A_i,P_A]=0\qquad (i=1,\ldots,N).
\end{equation}
Given that $A_i$
are genuinely incompatible, Lemma \ref{lemma4} implies that $P_A$
is proportional to $\mathbbm{1}_d$. This means that for any pure state
from $\mathrm{supp}(\rho_{AB})$ there exists a unitary $\mathcal{U}_B$ (different for a each state) such that 
\begin{equation}\label{SexyDuck}
    \ket{\psi}_{AB}=(\mathbbm{1}_d\otimes \mathcal{U}_B)\ket{\phi_+^d}.
\end{equation}

Having Eq. (\ref{SexyDuck}), let us now go back 
to the mixed state $\rho_{AB}$ and consider its eigendecomposition
\begin{equation}\label{Wedel}
    \rho_{AB}=\sum_{s=1}^Kp_s\proj{\psi_{s}}_{AB},
\end{equation}
where $\ket{\psi_s}_{AB}$ and $p_s\geq 0$ are the eigenvectors and eigenvalues of $\rho_{AB}$, respectively.
Clearly, every state in this decomposition admits the form 
(\ref{SexyDuck}), meaning that 
\begin{equation}\label{Schmidt2}
    \ket{\psi_s}_{AB}=\frac{1}{\sqrt{d}}\sum_{i=0}^{d-1}\ket{i}\ket{f_i^s}
\end{equation}
for some orthonormal bases $\{\ket{f_i^s}\}$ spanning the local supports of 
$\ket{\psi_s}_{AB}$, denoted $V_s=\mathrm{supp}(\Tr_A(\psi_s))$. The key observation 
is that the conditions (\ref{SOS00}) along with the facts that $A_i$ are genuine incompatible and that $\ket{\psi_s}_{AB}$
are eigenvectors of $\rho_{AB}$ force all these supports
$V_s$ to be pairwise orthogonal, which means that $\mathcal{H}_B=V_1\oplus\ldots\oplus V_K$.
Since $\dim V_s=d$, this implies that $\mathcal{H}_{B}=(\mathbbm{C}^d)_{B'}\otimes \mathcal{H}_{B''}$. Moreover, there exists a unitary $U_B$ such that 
\begin{equation}\label{Wawel}
    U_B\ket{f_i^s}=\ket{i}_{B'}\ket{s}_{B''}.
\end{equation}
Combining (\ref{Wedel}), (\ref{Schmidt}) and (\ref{Wawel})
we arrive at (\ref{lem1.2}) which together with Eq. (\ref{SOS00}) leads us to (\ref{lem1.1}). 
\end{proof}

An immediate consequence of Theorem \ref{Theo1M} is a simple method of
certification of projective measurements that are genuinely incompatible:
to certify that certain observables $A_i^*$ have been measured by Bob's untrusted device, it is enough to check whether an inequality (\ref{SteeIn}) constructed from these observables is maximally violated. In particular, any set of MUBs such as the one mentioned before can be certified in this way. To support this claim, in Lemma 2 in Appendix C of \cite{SupMat} we show that any pair of observables whose eigenbavses are mutually unbiased are GI.

{\it{Weaker form of self-testing.----}}Let us briefly discuss the case when the  observables on the trusted side $A_i$ are not GI, but still do not share a common eigenvector and can be used to violate the steering inequality \eqref{SteeIn} maximally. In such a case the steering inequality \eqref{SteeIn} 
can be used only for a partial certification of Bob's measurements.
%
%
To give an example let us consider the case $d=4$ and 
assume that Alice's observables are $A_1=Z_4$ and 
\be \label{ex}
A_2 = \sum^1_{j=0} (-1)^j \left( |+_j\ra\!\la +_j| +  \mathbbm{i}^{j+1} |-_j\ra\!\la -_j| \right),
\ee
where $|\pm_0\ra = (|0\ra \pm |1\ra)/\sqrt{2},$ $|\pm_1\ra = (|2\ra \pm |3\ra)/\sqrt{2}$. It can be checked that any matrix of the form
$P=\lambda_1\mathbbm{1}_2\oplus \lambda_2\mathbbm{1}_2$, where $\mathbbm{1}_2$
is the $2\times 2$ identity matrix and $\lambda_i\geq 0$ such that 
$\lambda_1^2+\lambda_2^2=1/2$,
commutes with both $A_i$. Thus, the maximal quantum value of the corresponding  inequality \eqref{SteeIn} is achieved by any state given by
%
%
$\lambda_1(|00\ra + |11\ra ) + \lambda_2 (|22\ra + |33\ra )$,
%
where, in particular, one of $\lambda_i$ might vanish. 
In other words, maximal violation of our inequality 
certifies an entangled subspace. However, 
%
%
whether $B_i$ can be certified depends on whether $P_A$ is full rank. For instance, if $\lambda_2=0$, then only the part of Bob's measurements which act on the subspace spanned by $\ket{0}$ and $\ket{1}$ can be certified; they are unitarily equivalent to $\Pi_2A^*_i\Pi_2$ where $\Pi_2=\ket{0}\!\bra{0}+\ket{1}\!\bra{1}$.  


{\it{Robustness.----}}In practical scenarios it becomes relevant to study robustness of our self-testing result against experimental imperfections such as the presence of noise in the preparation device or imperfect measurements. This problem for various combinations of measurements on the trusted side can be addressed numerically using the approach of \cite{Jordi}. However, finding an analytical statement for any $d$ is much more challenging. Here, assuming the underlying state to be pure and Bob's measurements to be projective, we address this issue and provide a way to find robustness bounds analytically for a class of genuinely $d-$outcome incompatible measurements representing MUBs of any dimension. 

\begin{thm}\label{Thm222}
Consider a pair of genuinely incompatible observables $A_1=X_dZ_d^{l}$ with $l=0,\ldots,d-1$ and $A_2=Z_d$, and assume that the corresponding steering inequality 
(considered already in Ref. \cite{Horodecki})
%
\begin{equation}\label{IneqRob}
\mathcal{B}_{d,2}:=\sum_{i=1}^2\sum_{k=0}^{d-1}\langle A_i^k\otimes B_i^k\rangle\leq \sqrt{2}(d-1)
\end{equation}
%
%
is violated by a state 
$\ket{\psi}_{AB}\in\mathbbm{C}^d\otimes\mathcal{H}_B$ 
and observables $B_i \ (i=1,2)$ acting on $\mathcal{H}_B$ such that $\mathcal{B}_{d,2}\geq 2(d-1) -\varepsilon$. Then, 
there exists a unitary $U_B:\mathcal{H}_B\to \mathcal{H}_B$ such that

\begin{eqnarray}\label{Rob1manu}
\left\|(\mathbbm{1}\otimes U_B)(\mathbbm{1}\otimes B_i^{k})\ket{\psi}-\mathbbm{1}\otimes (B'_i)^{k}\ket{\phi^+_d}\right\|\leq \sqrt{2\varepsilon}+2\sqrt{d}\sqrt[4]{2\varepsilon}\nonumber\\
\end{eqnarray}
and
\begin{eqnarray}\label{Rob2manu}
\left\|B_i^{k}-(B_i')^{k}\right\|_2\leq \sqrt{d}\,
\left(\sqrt{2\varepsilon}+4\sqrt{d}\sqrt[4]{2\varepsilon}\right),
\end{eqnarray}
where $k=0,\ldots,d-1$ and $B'_i$ are Bob's ideal observables, i.e., 
$B_i'=A_i^*$.
\end{thm}

\begin{proof}The proof is deferred to Appendix C of \cite{SupMat}.
\end{proof}

{\it{Conclusion.---}}Taking advantage of the steering scenario 
we propose a general method for certification
of a large class of $d$-outcome projective measurements
(including those corresponding to MUB),
termed here genuinely incompatible, and the 
maximally entangled state $\ket{\phi_{d}^{+}}$. Importantly, unlike the previous 
approaches, our scheme does not require assuming that the underlying 
state is pure and measurements are projective.

While the assumption that one of the measuring devices is trusted
makes our results less general as compared to the DI setting, 
they are still of relevance as far as 
certification of quantum systems is concerned. First, given a 
well-characterized reference measuring device, our scheme allows to
certify that any other device performs the same measurements
(of course, up the certain equivalences). Second, 
it can be applied to all 1SDI scenarios in which 
a client wants to certify the state provided by a provider along with its measuring device such as the quantum key distribution schemes \cite{Crypto1, Crypto2}. Third, from 
the experimental perspective it is easier to implement 1SDI protocols 
than DI protocols \cite{Crypto1}.  Last but not least, our results are 
interesting from the mathematical point of view as we introduce 
new techniques allowing to derive
certification schemes for arbitrary dimensional quantum systems.
It should also be stressed that, as pointed out recently in Ref. \cite{sarkar2},
in certain situations self-testing schemes derived within the steering scenario
can be made fully device-independent. At the same time, it is unclear whether such possibility exists for the prepare-and-measure certification schemes; in particular, the assumption that the underlying Hilbert space has bounded dimension can never be verified in a DI manner.

Some follow-up questions arise from our work. The most important one would be whether the certification method presented here can be made fully device-independent, for instance, by using the approach of \cite{Bowles, sarkar2, sarkar3}. Another interesting avenue would be to modify our scheme so that it allows to certify any set of genuinely incompatible measurements at the untrusted side by using the same fixed measurement at the trusted side such as $X_d$ and $Z_d$.


\begin{acknowledgements}
\textit{Acknowledgments.---}This work is supported by the National Science Center (Poland) through the SONATA BIS project No. 2019/34/E/ST2/00369.
\end{acknowledgements}

\input{ref7.bbl}



\onecolumngrid

\appendix

\section{Correlation form of steering functionals}\label{AppA}

Let us consider the standard form of the steering functional, i.e., 
\begin{equation}\label{SteeringFunctional}
    \sum_{b=0}^{d-1}\sum_{y=1}^N \Tr\left(F_{b|y}\sigma_{b|y}\right),
\end{equation}
where $\sigma_{b|y}$ form the assemblage and $F_{b|y}$ are some positive
semi-definite operators acting on $\mathbbm{C}^d$. Let us then consider a particular class of such functionals in which for every $y=1,\ldots,N$, the operators $F_{a|x}$ sum up to the identity, that is, 
\begin{equation}
    \sum_{a}F_{a|x}=\mathbbm{1},
\end{equation}
and therefore form a quantum measurement. With the aid of Fourier transform
we can represent each such a measurement by another set of $d$ operators 
\begin{equation}\label{Akx}
    A_{k|x}=\sum_{a=0}^{d-1}\omega^{ak} F_{a|x}
\end{equation}
with $k=0,\ldots,d-1$ and $\omega=\mathrm{exp}(2\pi\mathbbm{i}/d)$, which for the purposes of this article we simply call generalized observables; in fact if $F_{a|x}$ forms a projective measurement, then $A_{k|x}$ are unitary observables with outcomes labelled by the powers of $\omega$ (see also below). 

The inverse transform gives
\begin{equation}
F_{a|x}=\frac{1}{d}\sum_{k=0}^{d-1}\omega^{-ak} A_{k|x}.   
\end{equation}

Clearly, we can use the same approach to represent Bob's measurements $N_y=\{N_{b|y}\}_b$ in terms of generalized observables, that is, 
\begin{equation}\label{Bobs1}
    B_{k|y}=\sum_{b=0}^{d-1}\omega^{bk} N_{b|y},
\end{equation}
and
\begin{equation}\label{Bobs2}
    N_{b|y}=\frac{1}{d}\sum_{k=0}^{d-1}\omega^{-kb}B_{k|y}.
\end{equation}
It should also be noticed that the $B_{k|y}$ operators enjoy certain useful properties, that is, 
for any $k$ and $y$:
\begin{equation}\label{Bprop1}
    B_{k|y}^{\dagger}=B_{d-k|y}=B_{-k|y}
\end{equation}
and (see Proposition A.1 proven in Appendix A of Ref. \cite{Jed1})
\begin{equation}\label{Bprop2}
    B_{k|y}^{\dagger}B_{k|y}\leq \mathbbm{1}_B,
\end{equation}
where $\mathbbm{1}_B$ is the identity acting on $\mathcal{H}_B$, and finally $B_{0|y}=\mathbbm{1}$ for any $y$; clearly, the same identities are satisfied by $A_{k|x}$. Importantly, the $d$ operators $B_{k|y}$ fully determine Bob's measurements $N_y$. Slightly abusing the notation in what follows we denote $B_y=\{B_{k|y}\}_k$.

Let us now go back to the steering functionals (\ref{SteeringFunctional}). By exploiting the fact that $\sigma_{b|y}=\Tr_B[(\mathbbm{1}_A\otimes N_{b|y})\rho_{AB}]$, it can be rewritten as a combination of expectation values of the above generalized observables, 
\begin{equation}\label{SteeringFunctional2}
    \sum_{b=0}^{d-1}\sum_{y=1}^N \Tr\left(F_{b|y}\sigma_{b|y}\right)=\frac{1}{d}
   \sum_{k=0}^{d-1} \sum_{y=1}^N\left\langle A_{k|y}\otimes B_{-k|y}\right\rangle.
\end{equation}
We then further restrict the class of the steering functionals to the case when the operators $F_{a|x}$ form projective measurements for all $x$, meaning that they satisfy 
\begin{equation}
    F_{b|y}F_{b'|y}=\delta_{b,b'}F_{b|y}.
\end{equation}
It is not difficult to see that under the above assumption, $A_{k|y}$ are all unitary operators with eigenvalues $\omega^{bk}$ $(b=0,\ldots,d-1)$, and, moreover, $A_{k|y}$ for any $y$ are simply the $k$th powers of $A_{1|y}$. Using this fact, denoting
\begin{equation}
    A_{1|y}\equiv A_{y}, 
\end{equation}
dropping the $1/d$, and replacing $-k$ by $k$ which can be done with no loss of generality as
such a manipulation represents a simple relabelling the outcomes of Bob's measurements,
we finally arrive at our steering functionals of the form
\begin{equation}\label{SteeringFunctional21}
    \mathcal{B}_{d,N}=
   \sum_{k=1}^{d-1} \sum_{y=1}^N\left\langle A_{y}^k\otimes B_{k|y}\right\rangle,
\end{equation}
where we also got rid of the term for $k=0$ because 
the expectation value for this term is simply one.
It is worth recalling here that while the dimension of Alice's observables is known to be $d$, Bob's measurements act on a Hilbert space of in principle unknown but finite dimension.


Let us make two comments here. The first one is that if Bob's measurements are projective too, 
the $B_{k|y}$ operators are also unitary for all $k$ and $y$, and therefore $B_{k|y}$ in Eq. (\ref{SteeringFunctional21}) can be replaced by $B_{y}^k$.
%
%

Second, it should be stressed here that the form (\ref{SteeringFunctional21}) of the steering functional is particularly useful as far as self-testing is concerned because it is straightforward to find its maximal quantum value which simply equals its algebraic bound, i.e., $\beta_Q=N(d-1)$, as well as a quantum realization achieving this value. Precisely, 
such a realization consists of the maximally entangled state of two qudits 
\begin{equation}\label{MaxEntApp2}
    \ket{\phi_{+}^d}=\frac{1}{\sqrt{d}}\sum_{i}\ket{ii}
\end{equation}
and Bob's measurement satisfying $B_i=A_i^{*}$ for any $i$.

On the other hand, any quantum realization achieving $\beta_Q$ of our inequality for a given set of observables $A_i$, consisting of a mixed state $\rho_{AB}$ and Bob's measurements represented by $B_{i}$, 
must satisfy the following relations
\begin{equation}\label{Cabernet}
    A_y^k\otimes B_{k|y}\,\rho_{AB}=\rho_{AB}
\end{equation}
for any $y$ and $k$. This is because for maximal violation
every term in Eq. (\ref{SteeringFunctional21}) satisfies
\begin{equation}
\Tr[(A_y^k\otimes B_{k|y})\rho_{AB}]=1,
\end{equation}
which due to the facts that $A_y$ are unitary and that $B_{k|y}$ satisfy (\ref{Bprop2}), 
directly implies (\ref{Cabernet}).

\section{Bounding the maximal classical value of our functionals}
\label{AppB}

In this section we derive an upper bound on the maximal classical value
of the steering expression \eqnref{classBoun}, and show that for any set of 
genuinely $d$-outcome incompatible measurements the corresponding steering inequality is nontrivial, that is, $\beta_L<\beta_Q$. 
%

%
%


Let us begin by demonstrating how to derive the bound (\ref{classBoun}). To this end, we assume that the assemblage $\{\sigma_{b|y}\}$ admits a local hidden state model, that is, 
\begin{equation}
    \sigma_{b|y}=\sum_{\lambda}p(\lambda)p(b|y,\lambda)\rho_{\lambda},
\end{equation}
where $\lambda$ is some random variable with some probability distribution $p(\lambda)$, 
$p(b|y,\lambda)$ a probability distributions according to which Bob returns the outcomes $b$ given the measurement choice $y$ and the hidden variable $\lambda$, and $\rho_{\lambda}$
are some states acting on Alice's Hilbert space. In terms of the joint probability 
distributions such an LHS model expresses as
%
%
%
\begin{eqnarray}
p(a,b|x,y)=\sum_{\lambda}p(\lambda)p(a|x,\rho_\lambda)p(b|y,\lambda),
\end{eqnarray}
where $p(a|x,\rho_\lambda)$ is the probability of obtaining the outcome $a$
when measuring the observable $A_x$ on the state $\rho_{\lambda}$. Then, in terms of the expectation values appearing in (\ref{SteeringFunctional2}), 
the above rewrites as 
\begin{equation}
\left\langle A_{k|x}\otimes B_{l|y}\right\rangle_{\mathrm{LHS}}=
\sum_{\lambda}p(\lambda)\langle A_{k|x}\rangle_{\rho_{\lambda}} 
\langle B_{l|y}\rangle_{\lambda},
\end{equation}
%
%
where $\langle A_{k|x}\rangle_{\rho_{\lambda}}=\Tr[A_{k|x}\rho_{\lambda}]$ with $A_{k|x}$ being operators representing the $x$th measurement performed by Alice 
[cf. Eq. (\ref{Akx})]. Then, $\langle B_{l|y}\rangle_{\lambda}$ stands for the Fourier transform of $p(b|y,\lambda)$, i.e., 
\begin{equation}
   \langle B_{l|y}\rangle_{\lambda}=\sum_{b}\omega^{bl}p(b|y,\lambda).
\end{equation}

Now, our steering expression (\ref{SteeringFunctional21}) for LHS models reads
\begin{equation}
\langle\mathcal{\hat{B}}_d\rangle_{\mathrm{LHS}}=\sum_{\lambda}p(\lambda) \sum_{y=1}^N\sum_{k=1}^{d-1}\langle A_{k|y}\rangle_{\rho_{\lambda}} 
\langle B_{-k|y}\rangle_{\lambda},
\end{equation}
and by noting that $|\langle B_{l|y}\rangle_{\lambda}|\leq 1$ for any choice of $y$, $l$ and $\lambda$, its absolute value can be upper bounded as 
\begin{equation}
|\langle\mathcal{\hat{B}}_d\rangle_{\mathrm{LHS}}|\leq \sum_{\lambda}p(\lambda) \sum_{i=1}^N\sum_{k=1}^{d-1}|\langle A_i^k\rangle_{\rho_{\lambda}}|,
\end{equation}
and further as 
\begin{equation}\label{dupa}
|\langle\mathcal{\hat{B}}_d\rangle_{\mathrm{LHS}}|\leq \max_{\rho} \sum_{i=1}^N\sum_{k=1}^{d-1}|\langle A_i^k\rangle_{\rho}|,
\end{equation}
where we took advantage of the fact that the right-hand side of 
the above inequality is a convex combination and that by optimizing it 
over states $\rho$ acting on Alice's Hilbert space we cannot decrease 
its value. We thus obtain the bound (\ref{classBoun}).

A direct, but important for our considerations, corollary that stems from the above bound is that for LHS models with any set of genuinely incompatible measurements 
$A_i$, the right hand side of (\ref{SteeringFunctional21}) is smaller than $N(d-1)$, and thus $\beta_L<N(d-1)$, meaning that for any such a set our steering inequality is nontrivial.  
To see this explicitly let us assume that 
\begin{equation}
    \sum_{i=1}^N\sum_{k=0}^{d-1}|\langle A_i^k\rangle_{\rho}|=N(d-1),
\end{equation}
which implies that $|\langle A_i^k\rangle_{\rho}|=1$ for any $i$, $k$ and some $\ket{\psi}\in\mathbbm{C}^d$. Due to the fact that $A_i^k$ are unitary operators, this means that $\ket{\psi}$ is their eigenvector, possibly with different eigenvalues. Consequently, these observables share 
a common subspace spanned by this vectors which contradicts that they are genuinely incompatible.

\section{Self-testing}

Here we present proofs of the lemmas and theorems stated in the main text.
%

\label{AppC}

\subsection*{Proof of Lemma 1}

\setcounter{lem}{0}

\begin{lem}\label{lemma1}
Consider a set of $N$ $d$-outcome unitary observables $A_i$ obeying $A_i^d=\mathbbm{1}_d$ that are genuinely incompatible. If for some nonzero normal matrix $P$ acting on $\mathbbm{C}^d$, $[P,A_i]=0$ for every $i=1,\ldots,N$, then $P=\mu \mathbbm{1}$ with $\mathbbm{1}$ being the identity acting on $\mathbbm{C}^d$ and $\mu\in\mathbbm{C}$.
\end{lem}
\begin{proof}
The proof is by contradiction. Let us assume that the matrix $P$ is not proportional to the identity. Since it is normal, we can decompose it as
\begin{equation}
    P=\sum_{i}\lambda_i P_i,
\end{equation}
where $\lambda_i\in\mathbbm{C}$ are the eigenvalues of $P$ which without any loss of generality we can assume to be all different, and $P_i$ are the corresponding orthogonal eigenprojections. Here
we do not exclude the possibility that one of the eigenvalues is zero; below we show that 
this is in fact impossible and that $\mathrm{rank}(P)=d$.

Now, it is a known fact that if $[P,A_i]=0$, then $A_i$ is of block form 
\begin{equation}\label{dupa12}
 A_i=A_i^{(1)}\oplus \ldots \oplus A_i^{(m)},
\end{equation}
where each block $A_i^{(j)}$ acts on $\mathrm{supp}(P_j)$ and $m$ is the number of different eigenvalues of $P$. For completeness, let us add here a short proof of this fact. To this aim, we rewrite $[P,A_i]=0$ as
\begin{equation}
    A_iP=PA_i.
\end{equation}
We then sandwich this equation with $P_m$ and $P_n$ for $m\neq n$, which 
gives
\begin{equation}
    P_mA_iPP_n=P_mPA_iP_n.
\end{equation}
Noting that $PP_n=\lambda_nP_n$ and $P_mP=\lambda_m P_m$, 
we can rewrite the above as
\begin{equation}
    (\lambda_n-\lambda_m)P_mA_i P_n=0,
\end{equation}
and since $\lambda_n\neq \lambda_m$ for $n\neq m$, we conclude that 
$P_mA_i P_n=0$ for any pair $m\neq n$. Of course, this conclusion can be drawn
for each $A_i$, and thus we obtain (\ref{dupa12}). 

The fact that all observables can be written as in (\ref{dupa12}) implies that 
$\mathrm{supp}(P_i)$ is an invariant subspace of every $A_i$ of dimension lower than $d$, which contradicts the assumption. As a consequence, rank$(P)=d$. This completes the proof.
\end{proof}

\subsection*{Exact self-testing}

Let us now provide a full proof of Theorem 1 of the main text.
However, since the whole proof is very long, we split it into two 
parts. First, in Theorem 1.1 we prove our self-testing statement for pure states
and projective measurements on Bob's side. Then, Theorem 1.2 
extends this statement to the case of mixed states and nonprojective measurements 
on Bob's side. We stress here that in our considerations we can 
assume that the local state $\rho_B$ is full rank, or, in other words, 
that the Bob's measurement operators act on a Hilbert space of the same
dimension as $\mathcal{H}_B$. Let us also add that in what follows all
the considered observables are unitary and their eigenvalues are $\omega^i$
with $i=0,\ldots,d-1$.

\renewcommand{\thetheorem}{1.\arabic{theorem}}

\setcounter{thm}{0}

\begin{theorem}\label{Theo1} 
Assume that the steering inequality \eqref{SteeIn} is maximally violated by a state $\ket{\psi}_{AB}\in\mathbbm{C}^d\otimes\mathcal{H}_B$ and unitary $d$-outcome observables $A_i,B_i \ (i\in \{1,\ldots,N\})$ acting on, respectively, $\mathbbm{C}^d$ and $\mathcal{H}_B$ such that the observables $A_i$ on the trusted side are $d$-outcome genuinely incompatible. Then, the following statement holds true for any $d$: there exists a local unitary transformation on Bob side, $U_B:\mathcal{H}_B\rightarrow\mathcal{H}_B$, such that
\begin{eqnarray}\label{Theo1.2A}
(\mathbbm{1}_A\otimes U_B)\ket{\psi}_{AB}=
\ket{\phi_d^+}
\end{eqnarray}
and 
\begin{eqnarray}\label{Theo1.1A}
\forall i, \quad U_B\,\overline{B}_i\,U_B^{\dagger}=A_i^{*},
\end{eqnarray}
where $\overline{B}_i$ is $B_i$ projected onto the support of Bob's state $\rho_B$ acting on $\mathbbm{C}^d$.
\end{theorem}
\begin{proof}
The optimal quantum value $\beta_{Q}=N(d-1)$ of the steering inequality \eqref{SteeIn} is the same as its algebraic value. Thus, we have the following set of relations
[cf. Eq. (\ref{Cabernet})]
\begin{eqnarray}\label{SOSApp}
\quad  A_i^{k}\otimes B_i^{k} \ket{\psi}_{AB} = \ket{\psi}_{AB}
\end{eqnarray}
for all $i$ and $k$, where due to the fact that Bob's measurements are projective, 
$B_{k|y}=B_y^k$. For clarity the rest of the proof is divided into 
three main steps.\\

\noindent\textit{Step 1.} Let us first use these relations
to prove that the fact that for $A_i$ forming a set of
genuinely incompatible measurements implies that $\mathrm{rank}(\rho_A)=d$
with $\rho_A$ being the reduced density matrix corresponding to Alice, $\rho_A=\Tr_B(\proj{\psi}_{AB})$;
in other words, any state maximally violating our inequality 
for genuinely incompatible measurements must be of maximal local 
rank. To this end, we assume that $\mathrm{rank}(\rho_A)<d$ and consider (\ref{Cabernet}) 
for $k=1$ when $B_i$ is projective and $\rho_{AB}$ is pure, which implies that  
\begin{eqnarray}\label{SOS22}
\quad  \Pi_A\, A_i\,\Pi_A\otimes B_i \ket{\psi}_{AB} = \ket{\psi}_{AB},
\end{eqnarray}
where $\Pi_A$ is the projector onto the support of 
$\rho_A$. Denoting then $\overline{A}_i\equiv \Pi_A\, A_i\,\Pi_A$
and owing to the fact that $B_i$ are unitary and satisfy 
$B_i^d=\mathbbm{1}$, we conclude that $\overline{A}_i$ are unitary 
matrices on $\mathrm{supp}(\rho_A)$ and satisfy $\overline{A}_i^d=\Pi_A$.
It is then not difficult to see that since $A_i$ are unitary, they 
must be of block form 
\begin{equation}
    A_i=\overline{A}_i\oplus A'_i 
\end{equation}
for some square unitary matrix $A'_i$ of dimension $[d-\mathrm{rank}(\rho_A)]\times[d-\mathrm{rank}(\rho_A)]$.
However, this implies that all the observables $A_i$ have a
common invariant subspace of dimension lower $d$, contradicting the
assumption that they are genuinely incompatible on $\mathbbm{C}^d$. 
As a consequence, $\ket{\psi}_{AB}$ is of full local rank, or equivalently  $\mathrm{rank}(\rho_A)=d$ 
which also means that $\mathrm{rank}(\rho_B)=d$, where $\rho_B=\Tr_A(\proj{\psi}_{AB})$.\\

\noindent\textit{Step 2.} Using a similar reasoning we now show that Bob's observables 
all split into a direct sum 
\begin{equation}\label{block}
    B_i=\overline{B}_i\oplus E_i,
\end{equation}
where $\overline{B}_i=\Pi_B\,B_i\,\Pi_B$ are $d\times d$ matrices acting 
on the support of $\rho_B$, whereas $E_i$ are of unknown dimension and are defined outside $\mathrm{supp}(\rho_B)$; in particular, $E_i\ket{\psi}_{AB}=0$. Both $\overline{B}_i$ and 
$E_i$ are unitary and their spectra belong to $\{1,\ldots,\omega^{d-1}\}$. 

To prove the above statement, let us first notice that 
each $B_i$ can be written in a block form,
\begin{equation}
    B_i=\left(
    \begin{array}{cc}
    \overline{B}_i & C_i\\
    D_i & E_i
    \end{array}
    \right),
\end{equation}
where, as before, $\overline{B}_i=\Pi_B\,B_i\,\Pi_B$
and $E_i=\Pi_B^{\perp}\,B_i\,\Pi_B^{\perp}$
where $\Pi_B$ and $\Pi_B^{\perp}$ are the projectors onto the support of $\rho_B$ and its complement in $\mathcal{H}_B$, respectively. Let us then go back to Eq. (\ref{Cabernet}) for $k=1$ and project it onto the support of $\rho_B$, that is, 
\begin{equation}\label{Jumila}
 A_i\otimes \Pi_B\, B_i\,\Pi_B \ket{\psi}_{AB} \equiv A_i\otimes  \overline{B}_i\ket{\psi}_{AB} = \ket{\psi}_{AB}.
\end{equation}
By applying $A_i^{\dagger}\otimes \overline{B}_i^{\dagger}$ to this identity and using the fact that $A_i$ are unitary, we conclude that $\overline{B}_i^{\dagger}\overline{B}_i=\mathbbm{1}_d$. In a fully analogous way 
we obtain from Eq. (\ref{Cabernet}) for $k=d-1$ that $\overline{B}_i\overline{B}_i^{\dagger}=\mathbbm{1}_d$ and thus 
all $\overline{B}_i$ are unitary. On the other hand, by applying $A_i\otimes\overline{B}_i$ to Eq. (\ref{Jumila}) $d-1$ times, we
obtain that $\overline{B}_i^d=\mathbbm{1}_d$ and thus their spectra
belong to $\{1,\ldots,\omega^{d-1}\}$. 

It is then direct to see that since both $B_i$ and $\overline{B}_i$ 
are unitary on the respective Hilbert spaces, the off-diagonal blocks in Eq. (\ref{Jumila}) must vanish, $C_i=D_i=0$, leading to the desired block form 
(\ref{block}). This finally implies that $E_i$ must also be
unitary and their eigenvalues are powers of $\omega$. \\

\noindent\textit{Step 3.} Let us then consider the Schmidt decomposition of 
$\ket{\psi_{AB}}$,
\begin{equation}\label{SchmidtApp}
    \ket{\psi}_{AB}=\sum_{i=0}^{d-1}\lambda_i\ket{e_i}\ket{f_i},
\end{equation}
where due to the fact that $\mathrm{rank}(\rho_A)=d$, the Schmidt coefficient $\lambda_i$ $(i=0,\ldots,d-1)$ are all positive and satisfy $\sum_{i}\lambda_i^2=1$. Moreover, the local vectors $\ket{e_i}$ and $\ket{f_i}$ form orthonormal bases in $\mathbbm{C}^d$ and $\mathcal{H}_B$, respectively.

We can now make a few manipulations with (\ref{SchmidtApp}). 
First, we observe that there exists a unitary $U_B$ such that 
$\ket{f_i}=U_B\ket{e_i^*}$ for every $i$, where the asterisk denotes 
complex conjugation in the standard basis; thus
\begin{eqnarray}\label{state2App}
\ket{\psi}_{AB}=(P_A\otimes U_B)
\frac{1}{\sqrt{d}}\sum_{i=0}^{d-1}\ket{e_i}\ket{e_i^*},
\end{eqnarray}
where $P_A$ is a matrix diagonal in the $\{\ket{e_i}\}$ basis 
with eigenvalues $\sqrt{d}\,\lambda_i$; recall the fact that $A_i$ are GI implies that $P_A$ is full rank, $\mathrm{rank}(P_A)= d$. We finally observe that the state appearing on the right hand side of (\ref{state2App}) is the maximally entangled state (\ref{MaxEntApp2});  consequently %
\begin{equation}\label{PW}
\ket{\psi}_{AB}=(P_A\otimes U_B)\ket{\phi^{+}_d}.
\end{equation}

Now, combining Eqs. \eqref{PW} and \eqref{SOSApp} for $k=1$, we obtain
\begin{eqnarray}
(A_iP_A\otimes \widetilde{B}_i)\ket{\phi^{+}_d}=(P_A\otimes \mathbbm{1}_B)\ket{\phi^{+}_d},
\end{eqnarray}
where $\widetilde{B}_i=U_B\, \overline{B}_i\, U_B^\dagger$.
Using then the fact that $R\otimes Q\ket{\phi_d^{+}}=RQ^T\otimes\mathbbm{1}\ket{\phi_d^{+}}$
holds for any two $d\times d$ matrices $R$ and $Q$,
and omitting the state $\ket{\phi_d^+}$, we obtain 
a set of matrix equation 
\begin{equation}\label{CULO-A}
A_iP_A\widetilde{B}_i^T=P_A.
\end{equation}
Let us then consider the Hermitian conjugation of the above equation 
\begin{equation}\label{CULO-B}
\widetilde{B}_i^*P_A A_i^{\dagger}=P_A,
\end{equation}
where we used the fact that $P_A$ is self-adjoint. After multiplying 
Eq. (\ref{CULO-A}) by Eq. (\ref{CULO-B}) and using the fact that 
$\widetilde{B}_i$ are unitary, which implies that
$\widetilde{B}_i^T\widetilde{B}_i^*=(\widetilde{B}_i^{\dagger}\widetilde{B}_i)^*=\mathbbm{1}_d$, 
we arrive at
\begin{equation}
    A_i P_A^2A_i^{\dagger}=P_A^2.
\end{equation}
Since $A_i$ are unitary, the above implies a simple condition 
for the matrix $P_A$, $[A_i,P_A^2]=0$, which due to the fact that 
$P_A\geq0$ is equivalent to
\begin{equation}
    [A_i,P_A]=0
\end{equation}
for all $i$. Owing to the facts that $P_A\geq 0$ and that $A_i$
are genuinely incompatible, the above lemma tell us that $P_A$
is proportional to identity, proving Eq. (\ref{Theo1.2A}).
Then, plugging this particular form of $P_A$ into Eq. (\ref{CULO-A})
we obtain (\ref{Theo1.1A}), which completes the proof.
\end{proof}

Let us now proof the main result of our work, that is, 
Theorem 1 of the main text. We will extend the above self-testing 
statement of Theorem \ref{Theo1} to the case when the state shared 
by Alice and Bob is mixed which might for instance represent a cryptographic scenario with a malicious eavesdropper who keeps the purification $\ket{\psi_{ABE}}$ of $\rho_{AB}$.
Moreover, we do not assume Bob's measurements to be projective.
\begin{theorem}\label{Theo2}
Assume that for a given set of GI unitary observables $A_i$ obeying $A_i^d=\mathbbm{1}_d$, the steering inequality \eqref{SteeIn} is maximally violated by a state $\rho_{AB}$ acting on $\mathbbm{C}^d\otimes\mathcal{H}_B$ and measurements represented by $B_i$. Then, the following statements hold true for any $d$: (i) Bob's measurements are projective, that is, all operators $B_{k|y}$ are unitary such that $B_{k|y}^d=\mathbbm{1}$, (ii) Bob's Hilbert space decomposes as $\mathcal{H}_{B}=(\mathbbm{C}^d)_{B'}\otimes \mathcal{H}_{B''}$, and (iii) there exists a local unitary transformation on Bob side, $U_B:\mathcal{H}_B\rightarrow\mathcal{H}_B$, such that
\begin{eqnarray}\label{lem1.2}
(\mathbbm{1}_A\otimes U_B)\rho_{AB}(\mathbbm{1}_A\otimes U_B^{\dagger})=\proj{\phi^+_d}_{AB'}\otimes \rho_{B''}.
\end{eqnarray}
and
\begin{eqnarray}\label{lem1.1}
\forall i, \quad U_B\,B_i\,U_B^{\dagger}=A_i^{*}\otimes \mathbbm{1}_{B''},
\end{eqnarray}
where $B''$ denotes Bob's auxiliary system.
\end{theorem}
\begin{proof}
The departure point of the proof is the fact, already mentioned in Appendix \ref{AppA}, that 
if for a given set of observables $A_i$, our steering inequality is maximally violated by $\rho_{AB}$ and $B_{k|y}$, then
\begin{equation}\label{wazne}
  (A_i^{k}\otimes B_{k|i})\,\rho_{AB}=\rho_{AB}
\end{equation}
for every $i$ and $k$. \\

\noindent\textit{Step 1.} We begin by showing that Bob's measurements maximally violating our inequality are projective. Applying $A_i^{-k}\otimes B_{-k|i}$
to the above and exploiting the fact that $A_i^{-k}=(A_i^{k})^{\dagger}$
and $B_{-k|i}=B_{k|i}^{\dagger}$, and using the fact that $A_i$ are unitary, we
arrive at
\begin{equation}
  (\mathbbm{1}_A\otimes B_{k|i}^{\dagger }B_{k|i})\,\rho_{AB}=\rho_{AB},
\end{equation}
which directly implies that 
\begin{equation}
    B_{k|i}^{\dagger }B_{k|i}=\mathbbm{1}_{B},
\end{equation}
where $\mathbbm{1}_{B}$ is the identity on $\mathcal{H}_B$. This implies that $B_{k|i}$ are unitary for any choice of $i$ and $k$, which through Eq. (\ref{Bobs2}) means that all $N_{b|y}$ are orthogonal projections, and thus Bob's measurements are projective. In what follows we substitute $B_{k|i}=B_i^{k}$.\\

\noindent\textit{Step 2.} Let us now move on to proving the main result and consider a 
decomposition of $\rho_{AB}$ maximally violating our inequality into pure states
\begin{eqnarray}\label{mixed1}
\rho_{AB}=\sum_{s=1}^{K}p_s\proj{\psi_s}_{AB},
\end{eqnarray}
where $p_s\geq 0$ are such that $\sum_sp_s=1$ and $K$ is some positive integer. Clearly, we can assume that all $\ket{\psi_s}$ are linearly independent. In fact, with no loss of generality we can assume that the above is the eigendecomposition of $\rho_{AB}$, meaning that 
$\ket{\psi_s}$ are pairwise orthogonal, i.e., $\langle\psi_s|\psi_{s'}\rangle=\delta_{ss'}$.

Now, the fact that $\rho_{AB}$ achieves the maximal quantum value of our inequality implies that each pure state in the decomposition (\ref{mixed1}) achieves it too, and therefore any $\ket{\psi_s}_{AB}$ satisfies the following conditions 
\begin{equation}\label{D7}
    (A_i^k\otimes B_i^k)\ket{\psi_s}_{AB}=\ket{\psi_s}_{AB}
\end{equation}
for any $i$ and $k$. It follows from Theorem \ref{Theo1} that 
for any state satisfying such equations for genuinely incompatible 
observables $A_i$ there exists a unitary $U_B^s$ such that  
\begin{equation}\label{TintodeVerano}
    (\mathbbm{1}\otimes U_B^s)\ket{\psi_s}_{AB}=\ket{\phi_+^d}.
\end{equation}
It is worth stressing here that the unitary operations $U_B^s$
might be different for different $\ket{\psi_{s}}_{AB}$.

For further benefits we rewrite the above as 
\begin{equation}\label{TintodeVerano2}
    \ket{\psi_s}_{AB}=[\mathbbm{1}\otimes (U_B^s)^{\dagger}]\ket{\phi_+^d}=\frac{1}{\sqrt{d}}\sum_{i}\ket{i}\ket{f_i^s},
\end{equation}
where the vectors $\ket{f_i^s}=(U_B^s)^{\dagger}\ket{i}$ form an orthonormal basis
for any $s$. Moreover, 
\begin{equation}
    U_B^s\,\overline{B}_i^{(s)}\, (U_B^s)^{\dagger}=A_i^{*},
\end{equation}
where $\overline{B}_i^{s}$ denote Bob's observables $B_i$ projected onto
the local supports of $\ket{\psi_s}_{AB}$, that is, 
\begin{equation}
    \overline{B}_i^{(s)}=\Pi_B^s\, B_i\, \Pi_B^s,
\end{equation}
where $\Pi_B^s$ denotes the projector onto the support of $\rho_B^s=\Tr_A[\proj{\psi_s}_{AB}]$,
\begin{equation}
    \mathrm{supp}(\rho_B^s)\equiv V_s=\mathrm{span}\{\ket{f_0^s},\ldots,\ket{f_{d-1}^s}\}\subset \mathcal{H}_B.
\end{equation}
It is important to notice that as shown in the proof of Theorem \ref{Theo1}, the $d\times d$ matrices $\overline{B}_i^{(s)}$ are unitary and their spectrum belongs to the set $\{1,\omega,\ldots,\omega^{d-1}\}$. Moreover, from each such subspace Bob's observables $B_i$ decompose as (see the proof of Theorem \ref{Theo1})
\begin{equation}\label{Direct}
    B_i=\overline{B}_i^{(s)}\oplus E_i^{(s)},
\end{equation}
where $E_i^{(s)}$ is a unitary operator too that acts on the complement of $V_s$ in $\mathcal{H}_B$.

Our aim now is to prove that all the local subspaces $V_s$ corresponding to 
the pure states $\ket{\psi_{s}}_{AB}$ are mutually orthogonal. To this end, let us focus on the first two vectors $\ket{\psi_1}_{AB}$ and $\ket{\psi_2}_{AB}$ from the decomposition (\ref{mixed1}) and the corresponding subspaces $V_1$ and $V_2$. It will be very convenient for us to represent the maximally entangled state $\ket{\phi_+^d}$ in Eq. (\ref{TintodeVerano}) in the eigenbasis of one of Alice's observables, say $A_0$. Equivalently, due to the facts that $\ket{\phi_+^d}$ is invariant under the action of $U\otimes U^*$ for any unitary $U$ and that $U$ can be included on Bob's observables, without any loss of generality we can simply assume that the computational basis $\{\ket{i}\}$ is an eigenbasis of $A_0$, that is, 
\begin{equation}
 A_0\ket{i}=\omega^i\ket{i}\qquad (i=0,\ldots,d-1).   
\end{equation}
By applying this fact to Eq. (\ref{D7}) we see that 
both local bases $\{\ket{f_i^1}\}$ and $\{\ket{f_i^2}\}$ are in 
fact the eigenbases of $B_0$, that is
\begin{equation}\label{dupa}
    B_0\ket{f_i^s}=\omega^{-i}\ket{f_i^s}\qquad (s=1,2).
\end{equation}
This imposes certain orthogonality relations between vectors from 
both bases. Namely, due to the fact that $B_0$ is unitary, vectors that correspond to its 
different eigenvalues must be orthogonal, that is, 
\begin{equation}\label{Ortrelations}
    \langle f_i^1|f_j^2\rangle=0 \qquad (i\neq j).
\end{equation}
Thus, to prove that $V_1\perp V_2$ it's enough to show that 
the above orthogonality holds true also when $s=t$. In order to do so, 
we can always decompose
\begin{equation}\label{orthogonal}
    \ket{f_i^2}=\alpha_i\ket{f_i^1}+\beta_i\ket{g_i},
\end{equation}
where $\alpha_i,\beta_i\in\mathbbm{C}$ and $|\alpha_i|^2+|\beta_i|^2=1$
and $\ket{g_i}$ is a normalized vector such that $\langle f_i^1|g_i\rangle=0$ for any $i$.
Using then the orthogonality relations (\ref{Ortrelations})
and the fact that $\ket{f_i^1}$ form an orthonormal set, it is not difficult to 
see from Eq. (\ref{orthogonal}) that 
\begin{equation}\label{kolejne}
\beta_i\langle f_j^1|g_i\rangle=0\qquad (i,j=0,\ldots,d-1).
\end{equation}
This means that either $\beta_i=0$ or the vectors $\ket{g_i}$ for which $\beta_i\neq 0$ are orthogonal to the whole subspace $V_1$.

Let us now go back to the conditions (\ref{D7}) which for $k=1$ can be rewritten as
\begin{equation}
    (\mathbbm{1}_A\otimes B_i)\ket{\psi_s}_{AB}=(A_i^{\dagger}\otimes \mathbbm{1}_B)\ket{\psi_s}_{AB}.
\end{equation}
Using then the Schmidt decompositions of $\ket{\psi_s}_{AB}$ given in Eq. (\ref{TintodeVerano2}), 
the above gives
\begin{equation}
    \sum_{k=0}^{d-1}\ket{k}\otimes (B_i\ket{f_k^s})=\sum_{k=0}^{d-1}(A_i^{\dagger}\ket{k})\otimes \ket{f_k^s},
\end{equation}
which directly implies a set of vector equations that for the two considered vectors $\ket{\psi_1}_{AB}$ and $\ket{\psi_2}_{AB}$ we state explicitly as
\begin{equation}\label{Katarzynki1}
    B_i\ket{f_k^1}=\sum_{m=0}^{d-1}\langle k|A_i^{\dagger}|m\rangle\ket{f_m^1}
\end{equation}
and
\begin{equation}\label{Katarzynki2}
    B_i\ket{f_k^2}=\sum_{m=0}^{d-1}\langle k|A_i^{\dagger}|m\rangle\ket{f_m^2},
\end{equation}
where $k=0,\ldots,d-1$. We then use the decomposition 
(\ref{orthogonal}) in (\ref{Katarzynki2}), which 
leads us to 
\begin{equation}
\alpha_k B_i\ket{f_k^1}+\beta_kB_i\ket{g_k}=\sum_{m=0}^{d-1}\alpha_m\langle k|A_i^{\dagger}|m\rangle\ket{f_m^1}+\sum_{m=0}^{d-1}\beta_m\langle k|A_i^{\dagger}|m\rangle\ket{g_m}.
\end{equation}
Using then Eq. (\ref{Katarzynki1}) for $B_i\ket{f_i^1}$ and after some simple manipulations,
the above gives
\begin{equation}
\sum_{m=0}^{d-1}(\alpha_k-\alpha_m)\langle k|A_i^{\dagger}|m\rangle\ket{f_m^1}
=\sum_{m=0}^{d-1}\beta_m\langle k|A_i^{\dagger}|m\rangle\ket{g_m}-\beta_k B_i\ket{g_k}.
\end{equation}
If we sandwich the above equation with $\bra{f_n^1}$ we obtain
\begin{equation}
(\alpha_k-\alpha_n)\langle k|A_i^{\dagger}|n\rangle
=-\beta_k \bra{f_n^1}B_i\ket{g_k},
\end{equation}
where we used the fact that $\ket{f_n^1}\perp \ket{g_i}$ for any pair $n,i$ 
(see above for explanation). Using finally Eq. (\ref{kolejne}) together with the fact that $B_i$ decompose into blocks given in Eq. (\ref{Direct}), meaning that they act invariantly on the subspace spanned by $\ket{f_n^1}$, which can also inferred from Eq. (\ref{Katarzynki1}), it is direct to see that the right-hand side of the above equations simply vanishes, and hence,
\begin{equation}
(\alpha_k-\alpha_n)\langle k|A_i^{\dagger}|n\rangle=0\qquad (k,n=0,\ldots,d-1).
\end{equation}
The key observation now is that the above system of equations is satisfied 
if, and only if the diagonal matrix $Q=\mathrm{diag}[\alpha_0,\ldots,\alpha_{d-1}]$ 
commutes with every $A_i$, that is,
\begin{equation}
    [A_i^{\dagger},Q]=0
\end{equation}
for any $i$. However, as proven in Lemma \ref{lemma1} 
this is possible for genuinely incompatible observables $A_i$ iff 
$Q$ is proportional to the identity $\mathbbm{1}_d$. This means that all $\alpha_i$ are equal;
let us then denote $\alpha_i=\alpha$ for some $\alpha\in\mathbbm{C}$.

The final step is to use the fact that the vectors $\ket{\psi_{1}}_{AB}$
and $\ket{\psi_{2}}_{AB}$ are orthogonal which together with Eq. (\ref{Ortrelations})
gives
\begin{equation}
    0=\langle \psi_{1}|\psi_2\rangle=\frac{1}{d}\sum_{k}\langle f_k^1|f_k^2\rangle=\frac{1}{d}\sum_{k}\alpha_k=\alpha,
\end{equation}
and so $\alpha_i=\alpha=0$ for every $i$. If we plug this back to Eq. (\ref{orthogonal})
we see that $\ket{f_i^2}=\beta_i\ket{g_i}$ for some complex phase $\beta_i$. 
However, as already said $\ket{g_i}$ is orthogonal to $\ket{f_i^1}$ and hence
$\langle f_i^2|f_i^1\rangle=0$, which together with (\ref{Ortrelations})
implies finally that the subspaces $V_1$ and $V_2$ are orthogonal. 
Clearly, we can apply the same argumentation to every pair of subspaces $V_j$ and $V_k$, which allows us to conclude that they are all mutually orthogonal.\\

\noindent\textit{Step 3.} We 
are now ready to prove Eqs. (\ref{lem1.1}) and (\ref{lem1.2}). Indeed, the fact that the local supports $V_s$ are mutually orthogonal implies that Bob's Hilbert space admits the following decomposition
\begin{equation}
    \mathcal{H}_B= V_1\oplus V_2\oplus\ldots \oplus V_K,
\end{equation}
which given that $\dim V_s=d$, means in particular that $\mathcal{H}_B=(\mathbbm{C}^d)_{B'}\otimes\mathcal{H}_{B''}$ for some Hilbert space
$\mathcal{H}_{B''}$. Moreover, the fact that the vectors $\ket{f_{i}^s}$ span orthogonal 
subspaces for different $s$, implies that there exists a unitary operation $U_B:\mathcal{H}_B\to\mathcal{H}_B$ such that 
\begin{equation}
    U_B\ket{f_i^s}=\ket{i}_{B'}\otimes \ket{s}_{B''}
\end{equation}
with $i=0,\ldots,d-1$ and $s=1,\ldots,K$. Thus,
\begin{equation}\label{koniec2}
    (\mathbbm{1}_A\otimes U_B)\ket{\psi_s}_{AB}=\ket{\phi_+^d}_{AB'}\otimes\ket{s}_{B''}
\end{equation}
for every $s$, and therefore taking into account the decomposition 
(\ref{mixed1}) we finally have 
\begin{equation}\label{koniec}
    (\mathbbm{1}_A\otimes U_B)\rho_{AB}(\mathbbm{1}_A\otimes U_B^{\dagger})=\proj{\phi_+^d}\otimes\rho_{B''},
\end{equation}
where $\rho_{B''}=\sum_{s}p_s\proj{s}_{B''}$. This is precisely Eq. (\ref{lem1.2}) we wanted to prove. 

Having Eq. (\ref{koniec}) we can finally establish Eq. (\ref{lem1.1}). To this end, 
we notice that since $\mathcal{H}_{B}=(\mathbbm{C}^d)_{B'}\otimes \mathcal{H}_{B''}$, 
we can $U_B\,B_i\,U_B^{\dagger}$ in the following block form
\begin{equation}\label{Biblock}
 U_B\,B_i\,U_B^{\dagger}=\sum_{s,t=1}^K B_{s,t}^{i}\otimes\ket{s}\!\bra{t}_{B''},
\end{equation}
where $B_{s,t}^i$ are $d\times d$ blocks acting on $(\mathbbm{C}^d)_{B'}$, whereas 
$\ket{s}$ is the standard basis of $\mathcal{H}_{B''}$ and simultaneously the eigenbasis
of $\rho_{B''}$. Plugging Eqs. (\ref{koniec}) and (\ref{Biblock}) into Eq. (\ref{wazne}) for $k=1$,
we obtain the following matrix equation
\begin{equation}\label{blocks}
    \sum_{s,t}(A_i\otimes B_{s,t}^i)\proj{\phi_+^d}\otimes p_t\ket{s}\!\bra{t}_{B''}=\proj{\phi_+^d}\otimes
    \sum_{s}p_s\proj{s}_{B''}.
\end{equation}
The off-diagonal terms on $B''$ subsystem imply $(A_i\otimes B_{s,t}^i)\ket{\phi_+^d}=0$, 
which, given that $A_i$ are unitary, directly give $B_{s,t}^i=0$ for $s\neq t$. Then, 
the diagonal terms of (\ref{blocks}) lead us to 
\begin{equation}
    (A_i\otimes B_{s,s}^i)\ket{\phi_+^d}=\ket{\phi_{+}^d},
\end{equation}
which by virtue of $R\otimes Q\ket{\phi_d^{+}}=RQ^T\otimes\mathbbm{1}\ket{\phi_d^{+}}$, which holds true for any two matrices $R$ and $Q$, is equivalent to $B_{s,s}^i=A_i^{*}$. 
Substituting this back to Eq. (\ref{Biblock}) one obtains
\begin{equation}
     U_B\,B_i\,U_B^{\dagger}=\sum_{s=1}^K A_{i}^*\otimes\proj{s}_{B''}=A_{i}^*\otimes\mathbbm{1}_{B''}, 
\end{equation}
which is exactly Eq. (\ref{lem1.1}). This completes the proof.
\end{proof}

Now, we move on to prove that any set of observables whose eigenvectors
form mutually unbiased bases are genuinely incompatible. 
%
\begin{lem}\label{Lem2A}
Any two $d$-outcome observables whose eigenbases are mutually unbiased are genuinely incompatible.
\end{lem}
\begin{proof}
The proof is by contradiction. Let us begin by taking two $d-$dimensional mutually unbiased bases denoted by, $\{\ket{s_i}\}$ and $\{\ket{t_j}\}$ such that $|\langle s_i|t_j\rangle|^2=1/d$ for all $i,j\in\{0,1,\ldots,d-1\}$. Let us then construct observables from these bases, 
\begin{equation}
A_1=\sum_{i=0}^{d-1}\omega^{f(i)}\ket{s_i}\!\bra{s_i},\qquad  A_2=\sum_{i=0}^{d-1}\omega^{g(i)}\ket{t_i}\!\bra{t_i},
\end{equation}
where $f$ and $g$ are some permutations of the $d$-element set $\{0,\ldots,d-1\}$,
and assume that they share a common invariant subspace $V\subsetneq \mathbbm{C}^d$ such that $\dim V<d$. This means that both can be written as 
$A_i=A_i'\oplus A_i''$, where $A_i'$ acts on $V$, whereas 
$A_i''$ on its complement in $\mathbbm{C}^d$. Then, $A_1$ has eigenvectors
that are orthogonal to eigenvectors of $A_2$, which contradicts that 
the eigenvectors of $A_1$ and $A_2$ are mutually unbiased.
\end{proof}

\subsection*{Examplary set of genuinely incompatible observables which are not pairwise genuinely incompatible}

For a note, we construct a set of genuinely incompatible measurements for $d=4$ which are not pairwise genuinely incompatible. For this, we would require the underlying lemma.
\begin{lem}\label{lemma3}
Two invertible and diagonalizable matrices $T$ and $T'$ of dimension $D$, share a common non-trivial invariant subspace of dimension $d<D$ iff there exist $d$ eigenvectors of $T'$ which can be written as a linear combination of $d$ eigenvectors of $T$.
\end{lem}  

\begin{proof}
Let us first recall that any non-trivial invariant subspace of a matrix $T$ is spanned by its eigenvectors which implies that if $V_d$ is a $d-$dimensional common invariant subspace of $T$ and $T'$, then $V_d$ is spanned by $d$ eigenvectors of $T$ as well as $d$ eigenvectors of $T'$. Since, eigenvectors of any matrix are linearly independent, this implies that there exist $d$ eigenvectors of $T'$ which can be written as linear combinations of $d$ eigenvectors of $T$. 

Also, if $d$ eigenvectors of $T'$, denoted by $\{\ket{e'_i}\}|_{i=1,\ldots,d}$ can be written as a linear combination of $d$ eigenvectors of $T$, denoted by $\{\ket{e_i}\}|_{i=1,\ldots,d}$, then there exist a common invariant subspace $V_d$ of $T$ and $T'$ which is spanned by $\{\ket{e'_i}\}|_{i=1,\ldots,d}$. This completes the proof.
\end{proof}  
The example is stated below.
\begin{small}
\begin{eqnarray}\label{C19}
    A_1=\frac{1}{2}\begin{pmatrix}
1+\mathbbm{i} & 1-\mathbbm{i} & 0 & 0\\
1-\mathbbm{i} & 1+\mathbbm{i} & 0 & 0\\
0 & 0 & -2 & 0\\
0 & 0 & 0 & -2\mathbbm{i}
\end{pmatrix}, \quad   
A_2=\frac{1}{2}\begin{pmatrix}
2 & 0 & 0 & 0\\0 & -1+\mathbbm{i} & 1+\mathbbm{i} & 0\\0 &
1+\mathbbm{i}& -1+\mathbbm{i} & 0\\
0 & 0 & 0 & -2\mathbbm{i}
\end{pmatrix}, \quad
 A_3=\frac{1}{2}\begin{pmatrix}
2 & 0 & 0 & 0\\0 & 2\mathbbm{i} & 0 & 0\\0 & 0 &-1-\mathbbm{i} & \mathbbm{i}-1 \\0 &
0 & \mathbbm{i}-1& -1-\mathbbm{i} \end{pmatrix}
\end{eqnarray}
\end{small}

As discussed before, any non-trivial invariant subspace of a matrix $T$ is spanned by its eigenvectors of $T$. Using this fact, any non-trivial invariant subspace of $A_1$ is spanned by the vectors $\{\ket{+_{01}},\ket{-_{01}},\ket{2},\ket{3}\}$, any non-trivial invariant subspace of $A_2$ is spanned by the vectors $\{\ket{0},\ket{+_{12}},\ket{-_{12}},\ket{3}\}$ 
and any non-trivial invariant subspace of $A_3$ is spanned by the vectors $\{\ket{0},\ket{1},\ket{+_{23}},\ket{-_{23}}\}$ where $|\pm_{ij}\ra = (|i\ra \pm |j\ra)/\sqrt{2}$.

Using Lemma \ref{lemma3}, we can conclude that $A_1$ and $A_3$ have two common nontrivial invariant subspaces, spanned by $\{\ket{0},\ket{1}\}$ and $\{\ket{2},\ket{3}\}$. However, these two are not invariant subspaces of $A_2$ because it is not possible to represent an arbitrary vector from those subspaces as linear combinations of two of the eigenvectors of $A_2$. Thus, we can conclude that there is no common invariant subspace shared between $A_1, A_2$ and $A_3$.


\subsection*{Robust self-testing}

Finally, we prove that for some particular choice of Alice's observables, that is,
$A_1=X_dZ_d^{l}$ with $l=0,\ldots,d-1$ and $A_2=Z_d$,
our self-testing statements are robust to noises or experimental imperfections. 
However, for simplicity we assume here that the underlying state is pure and Bob's observables
are projective. 

For the above particular choice of Alice's measurements the corresponding steering inequality reads
\begin{equation}
\mathcal{B}_{d,2}:=\sum_{i=1}^2\sum_{k=0}^{d-1}\langle A_i^k\otimes B_i^k\rangle\leq \sqrt{2}(d-1)
\end{equation}
and was already considered in Ref. \cite{Horodecki}, where its classical bound was found to be $\beta_L=\sqrt{2}(d-1)$.

\setcounter{thm}{1}

\begin{thm}
Consider a pair of genuinely incompatible observables $A_1=X_dZ_d^{l}$ with $l=0,\ldots,d-1$ and $A_2=Z_d$, and assume that the corresponding steering inequality 
%
\begin{equation}
\mathcal{B}_{d,2}:=\sum_{i=1}^2\sum_{k=0}^{d-1}\langle A_i^k\otimes B_i^k\rangle\leq \sqrt{2}(d-1)
\end{equation}
%
%
is violated by a state 
$\ket{\psi}_{AB}\in\mathbbm{C}^d\otimes\mathcal{H}_B$ 
and observables $B_i \ (i=1,2)$ acting on $\mathcal{H}_B$ such that $\mathcal{B}_{d,2}\geq 2(d-1) -\varepsilon$. Then, 
there exists a unitary operation $U_B:\mathcal{H}_B\to \mathcal{H}_B$ such that

\begin{eqnarray}\label{Rob1}
\left\|(\mathbbm{1}\otimes U_B)(\mathbbm{1}\otimes B_i^{k})\ket{\psi}-\mathbbm{1}\otimes (B'_i)^{k}\ket{\phi^+_d}\right\|\leq \sqrt{2\varepsilon}+2\sqrt{d}\sqrt[4]{2\varepsilon}
\end{eqnarray}
and
\begin{eqnarray}\label{Rob2}
\left\|B_i^{k}-(B_i')^{k}\right\|_2\leq \sqrt{d}\,
\left(\sqrt{2\varepsilon}+4\sqrt{d}\sqrt[4]{2\varepsilon}\right),
\end{eqnarray}
where $k=0,\ldots,d-1$ and $B'_i$ are Bob's ideal observables, i.e., 
$B_i'=A_i^*$, and $\|\cdot\|_2$ stands for the Hilbert-Schmidt norm.
\end{thm}
\begin{proof}The first step in proving the above statements 
is to observe that 
$\mathcal{B}_{d,2}\geq 2(d-1) -\varepsilon$ implies the following set of inequalities
\begin{eqnarray}\label{App:1}
\left|\langle\psi| A_i^k\otimes B_i^k|\psi\rangle\right|\geq \mathrm{Re}\left(\langle\psi| A_i^k\otimes B_i^k|\psi\rangle\right)\geq 1-\varepsilon
\end{eqnarray}
for any $i=1,2$ and $k=1,\ldots,d-1$. 

The second observation we will exploit in our proof is that the shared state $\ket{\psi}$ can always be written in the basis of $Z_d$ as
\begin{eqnarray}\label{stateRob}
\ket{\psi}=\sum_{i=0}^{d-1}\alpha_i\ket{i}\ket{b_i},
\end{eqnarray}
where $\ket{b_i}$ are some, not necessarily orthogonal vectors from $\mathcal{H}_B$, and $\alpha_i$ are non-negative numbers satisfying $\sum_{i=0}^{d-1}\alpha_i^2=1$.

Let us first exploit Ineq. (\ref{App:1}) for $i=1$. Exploiting the fact that 
$(X_dZ_d^l)^{k}\ket{i}=\omega^{kl\left(i+\frac{k-1}{2}\right)}\ket{i+k}$, where the addition is mod $d$, we obtain
\begin{equation}
\sum_{i}\alpha_i\alpha_{i+k}\mathrm{Re}\left(\omega^{kl\left(i+\frac{k-1}{2}\right)}\langle b_{i+k}|B_1^k|b_i\rangle\right)\geq 1-\varepsilon
\end{equation}
which by virtue of the facts that $\mathrm{Re}(z)\leq |z|$
for any $z\in\mathbbm{C}$ and that $\left|\omega^{kl\left(i+\frac{k-1}{2}\right)}\langle b_{i+k}|B_1^k|b_i\rangle\right|\leq 1$ because $B_1$ is unitary and $\ket{b_i}$ are normalized, implies
\begin{equation}
\sum_{i}\alpha_i\alpha_{i+k}\geq 1-\varepsilon
\end{equation}
for any $k=0,\ldots,d-1$. By summing the above relation over $k$, one obtains
\begin{equation}\label{App:3}
\sum_{i,k=0}^{d-1}\alpha_i\alpha_{i+k}=\left(\sum_{i=0}^{d-1}\alpha_{i}\right)^2\geq d(1-\varepsilon)
\end{equation}
which gives 
\begin{equation}\label{App:2}
\sum_{i=0}^{d-1}\alpha_{i}\geq \sqrt{d}\sqrt{1-\varepsilon}.
\end{equation}
We can use this inequality to show that each $\alpha_i$ is 
"$\varepsilon$-close" to 
$1/\sqrt{d}$. To this end, let us write that
\begin{eqnarray}
    \sum_{i=0}^{d-1}\left(\alpha_i-\frac{1}{\sqrt{d}}\right)^2&=&\sum_{i=0}^{d-1}
    \alpha_i^2-\frac{2}{\sqrt{d}}\sum_{i=0}^{d-1}\alpha_i+1\nonumber\\
    &=&2\left(1-\frac{1}{\sqrt{d}}\sum_{i=0}^{d-1}\alpha_i\right),
\end{eqnarray}
where we used that $\sum_{i}\alpha_i^2=1$. Now, by using (\ref{App:2})
we can upper bound the above expression as
\begin{eqnarray}
    \sum_{i=0}^{d-1}\left(\alpha_i-\frac{1}{\sqrt{d}}\right)^2&\leq &
    2(1-\sqrt{1-\varepsilon})\nonumber\\
    &\leq& 2\varepsilon,
\end{eqnarray}
where we used $1-\sqrt{1-\varepsilon}\leq \varepsilon$, which in turn is a simple consequence of the fact that $\sqrt{1-\varepsilon}\geq 1-\varepsilon$ for any $0\leq\varepsilon\leq1$. This finally implies that for any $i$,
\begin{equation}\label{App:4.1}
\frac{1}{\sqrt{d}}-\sqrt{2\varepsilon}\leq \alpha_i\leq 
\frac{1}{\sqrt{d}}+\sqrt{2\varepsilon}.
\end{equation}

In a similar manner we can prove that 
\begin{equation}\label{App:4}
\frac{1}{d}-\sqrt{2\varepsilon}\leq \alpha_i\alpha_{i+j}\leq 
\frac{1}{d}+\sqrt{2\varepsilon}.
    \end{equation}
for any $i,j=0,\ldots,d-1$. Precisely, let us consider the following expression
\begin{eqnarray}
    \sum_{i,j=0}^{d-1}\left(\alpha_{i}\alpha_{i+j}-\frac{1}{d}\right)^2&=&2-\frac{2}{d}\sum_{i,j=0}^{d-1}\alpha_i\alpha_{i+j}\nonumber\\
    &\leq &2-\frac{2}{d}d(1-\varepsilon)=2\varepsilon,
\end{eqnarray}
where we have used Eq. (\ref{App:3}). By noting that each term in the above sum must also be upper bounded by $2\varepsilon$ we arrive at (\ref{App:4}).

Let us now use conditions (\ref{App:1}) for $A_2=Z_{d}$, which by using the fact that $Z_d^k\ket{i}=\omega^{ik}\ket{i}$ can be rewritten 
as
\begin{equation}
    \sum_{i=0}^{d-1}\alpha_i^2\,\mathrm{Re}\left(\omega^{ik}\langle b_i|B_2^k|b_i\rangle\right)\geq 1-\varepsilon.
\end{equation}
Using then the fact that $\mathrm{Re}(\omega^{ik}\langle b_i|B_2^k|b_i\rangle)\leq 1$ for any $i,k$, we have
\begin{equation}
    \sum_{i\neq j}\alpha_i^2+\alpha_j^2\,\mathrm{Re}\left(\omega^{jk}\langle b_j|B_2^k|b_j\rangle\right)\geq 1-\varepsilon,
\end{equation}
which by virtue of $\sum_i\alpha_i^2=1$ implies
\begin{equation}
    \alpha_j^2\left[1-\mathrm{Re}\left(\omega^{jk}\langle b_j|B_2^k|b_j\rangle\right)\right]\leq \varepsilon.
\end{equation}
Using then the left-hand side inequality of Eq. (\ref{App:4.1}) 
and the fact the term in the square brackets is non-negative, we obtain
\begin{equation}
    \frac{1}{d}\left[1-\mathrm{Re}\left(\omega^{jk}\langle b_j|B_2^k|b_j\rangle\right)\right]\leq \varepsilon+\sqrt{2\varepsilon}\leq 2\sqrt{2\varepsilon},
\end{equation}
which, by rearranging terms gives
\begin{equation}\label{App:5}
\mathrm{Re}\left(\omega^{jk}\langle b_j|B_2^k|b_j\rangle\right)\geq 1-2d\sqrt{2\varepsilon}.
\end{equation}
Let us now consider the eigendecompositon of $B_2$,
\begin{equation}
    B_2=\sum_{i=0}^{d-1}\omega^i P_i,
\end{equation}
where in general the orthogonal projections $P_i$ may be of higher rank than one.
By plugging it into (\ref{App:5}) one arrives at
\begin{equation}
    \sum_{i=0}^{d-1}\mathrm{Re}\left(\omega^{(i+j)k}\langle b_j|P_i|b_j\rangle\right)\geq 1-2d\sqrt{2\varepsilon},
\end{equation}
which after summing over $k=0,\ldots,d-1$ gives 
\begin{equation}\label{App:6}
    \langle b_j|P_{-j}|b_j\rangle\geq 1-2d\sqrt{2\varepsilon}.
\end{equation}
This inequality means that each vector $\ket{b_j}$ is $\varepsilon$-close to 
the subspace corresponding to the $\omega^{-j}$ outcome of $B_2$. Let us then 
introduce vectors $\ket{v_j}=P_{-j}\ket{b_j}$ and their normalized versions
$\ket{\overline{v}_j}=\ket{v_j}/\|\ket{v_j}\|$. From the fact that the projections $P_j$ are pairwise orthogonal we infer that $\ket{\overline{v}_j}$ are orthogonal as well. Moreover, Eq. (\ref{App:6}) implies that $\|\ket{v}_j\|\geq 1-2d\sqrt{2\varepsilon}$, and so the norm of each $\ket{v_j}$ is close to unity.

Now, due to the fact that by the very construction each $\ket{\overline{v}_j}$
belongs to the subspace corresponding to $P_{-j}$, the observable $B_2$ can be written as
\begin{equation}
    B_2=\sum_{i=0}^{d-1}\omega^{-i}\proj{\overline{v}_i}\oplus B_2',
\end{equation}
where $B_2'$ is an operator whose support is orthogonal to 
all $\ket{v_i}$. Then, due to the fact that $\ket{v_i}$ are orthogonal
one can find a unitary operation such that $U_B\ket{\overline{v}_i}=\ket{d-i}$. This operation brings $B_2$ to
\begin{equation}\label{App:65}
    U_B B_2 U_B^{\dagger}=\sum_{i=0}^{d-1}\omega^i\proj{i}\oplus B_2''.
\end{equation}
Denoting then $\ket{b_i'}=U_B\ket{b_i}$, where $\ket{b_i}$ are vectors appearing in the decomposition of $\ket{\psi}$ in Eq. (\ref{stateRob}), we deduce from
Eq. (\ref{App:6}) that
\begin{equation}\label{App:7}
    \langle b_i'|i\rangle\geq 1-2d\sqrt{2\varepsilon}.
\end{equation}

Given conditions (\ref{App:7}) we can now move on to proving 
our main statements (\ref{Rob1}) and (\ref{Rob2}). To this end, 
we notice that for the unitary operation $U_B$ introduced above, one has
\begin{eqnarray}\label{App:75}
\left\|(\mathbbm{1}\otimes U_BB_i^k)\ket{\psi}-[\mathbbm{1}\otimes (B'_i)^{k}]\ket{\phi^+_d}\right\|&=&
\left\|(A_i^{k}\otimes U_B B_i^{k})\ket{\psi}-[A_i^{k}\otimes (B'_i)^{k}]\ket{\phi^+_d}\right\|\nonumber\\
&=&\left\|(A_i^{k}\otimes \widetilde{B}_i^{k})\ket{\widetilde{\psi}}-\ket{\widetilde{\psi}}+\ket{\widetilde{\psi}}-\ket{\phi^+_d}\right\|,
\end{eqnarray}
where the first equation is a consequence of the fact that 
$A_i^k$ are unitary and that the vector norm is unitarily invariant, whereas the second equation has been obtained by adding and subtracting $\ket{\widetilde{\psi}}$ with the latter defined as $\ket{\widetilde{\psi}}=(\mathbbm{1}\otimes U_B)\ket{\psi}$ and 
$\widetilde{B}_i=U_B B_iU_B^{\dagger}$; in particular $\widetilde{B}_2$ is given in Eq. (\ref{App:65}). 

Using then the triangle inequality we arrive at
\begin{eqnarray}\label{App:8}
\left\|(A_i^{k}\otimes \widetilde{B}_i^{k})\ket{\widetilde{\psi}}-\ket{\widetilde{\psi}}+\ket{\widetilde{\psi}}-\ket{\phi^+_d}\right\|\leq\left\|(A_i^{k}\otimes \widetilde{B}_i^{k})\ket{\widetilde{\psi}}-\ket{\widetilde{\psi}}\right\|+\left\|\ket{\widetilde{\psi}}-\ket{\phi^+_d}\right\|
\end{eqnarray}
Let us now consider the first term of the right-hand side of the above inequality
and write is as
\begin{eqnarray}\label{App:85}
    \left\|(A_i^{k}\otimes \widetilde{B}_i^{k})\ket{\widetilde{\psi}}-\ket{\widetilde{\psi}}\right\|&=&\left\|(A_i^{k}\otimes B_i^{k})\ket{\psi}-\ket{\psi}\right\|\nonumber\\
    &=&\left\{2\left[1-\mathrm{Re}\left(\langle \psi|A_i^{k}\otimes B_i^{k}|\psi\rangle\right)\right]\right\}^{1/2}\nonumber\\
    &\leq &\sqrt{2\varepsilon},
\end{eqnarray}
where to get the inequality we have used conditions (\ref{App:1}). 

Let us then consider the second term appearing on the right-hand side of 
Eq. (\ref{App:8}) and rewrite it as
\begin{eqnarray}\label{App:10}
    \left\|\ket{\widetilde{\psi}}-\ket{\phi^+_d}\right\|&=&\left\{2\left[1-\mathrm{Re}(\langle\widetilde{\psi}|\phi_{d}^+\rangle)\right]\right\}^{1/2}\nonumber\\
    &=&\left\{2\left[1-\frac{1}{\sqrt{d}}\sum_{i}\alpha_i\,\mathrm{Re}
    (\langle\overline{b}_i|i\rangle)\right]\right\}^{1/2}\nonumber\\
    &\leq &\left\{2\left[1-\frac{1}{\sqrt{d}}(1-2d\sqrt{2\varepsilon})\sum_{i}\alpha_i\right]\right\}^{1/2}\nonumber\\
    &\leq & \left\{2\left[1-(1-2d\sqrt{2\varepsilon})\sqrt{1-
    \varepsilon}\right]\right\}^{1/2}\nonumber\\
    &\leq & 2\sqrt{d}\sqrt[4]{2\varepsilon},
\end{eqnarray}
where to obtain the first inequality we have used Eqs. (\ref{App:7})
and (\ref{App:2}), respectively, whereas the third inequality stems from 
two simple inequalities: $1-\sqrt{1-\varepsilon}\leq \varepsilon$ and $\varepsilon\leq \sqrt{2\varepsilon}$. By applying Eqs. (\ref{App:8}), (\ref{App:85}) and (\ref{App:10}) into (\ref{App:75}) we obtain (\ref{Rob1}).

Let us eventually prove (\ref{Rob2}). For this purpose, we first observe that
\begin{equation}\label{App:9}
  \left\|\widetilde{B}_i^{k}-(B'_i)^{k}\right\|_2^2=d\left\|\left[\widetilde{B}_i^{k}-(B'_i)^{k}\right]\ket{\phi_d^{+}}\right\|^2.
\end{equation}
Then, the triangle inequality gives us
\begin{eqnarray}
    \left\|\widetilde{B}_i^{k}\ket{\widetilde{\psi}}-(B'_i)^{k}\ket{\phi_d^+}\right\|&=&
    \left\|\widetilde{B}_i^{k}\ket{\widetilde{\psi}}-(B'_i)^{k}\ket{\phi_d^+}+(B'_i)^k\ket{\phi_d^+}-\widetilde{B}_i^k\ket{\phi_d^+}\right\|\nonumber\\
    &\geq&\left\|\left[\widetilde{B}_i^{k}-(B'_i)^k\right]\ket{\phi_d^+}\right\|-\left\|\widetilde{B}_i^{k}(\ket{\widetilde{\psi}}-\ket{\phi_d^+})\right\|,
\end{eqnarray}
which by taking into account the identity (\ref{App:9}) and the unitary invariance of the vector norm, leads us to
\begin{equation}
    \left\|\widetilde{B}_i^{k}-(B'_i)^{k}\right\|_2\leq  \sqrt{d}\left(\left\|\widetilde{B}_i^{k}\ket{\widetilde{\psi}}-(B'_i)^{k}\ket{\phi_d^+}\right\|+\left\|\ket{\widetilde{\psi}}-\ket{\phi_d^+}\right\|\right).
\end{equation}
Using inequalities (\ref{Rob1}) and (\ref{App:10}) we obtain 
(\ref{Rob2}).
\end{proof}

\end{document}

%% file: ref7.bbl
\providecommand{\noopsort}[1]{}\providecommand{\singleletter}[1]{#1}%
%